\documentclass[%
 reprint,
 amsmath,amssymb,
 aps,
showkeys,
]{revtex4-2}

\usepackage{graphicx}
\usepackage{dcolumn}
\usepackage{bm}

\begin{document}


\title{Reservoir computing based on a silicon microring and time multiplexing for binary and analog operations}

\author{Massimo Borghi}
 \email{corresponding author: massimo.borghi@unitn.it}
\author{Stefano Biasi}%
\author{Lorenzo Pavesi}
\affiliation{Nanoscience laboratory, Department of Physics, University of Trento, Via Sommarive 14, 38123, Trento,
Italy
}%




\date{\today}

\begin{abstract}
Photonic implementations of reservoir computing (RC) promise to reach ultra-high bandwidth of operation with moderate training efforts. Several optoelectronic demonstrations reported state of the art performances for hard tasks as speech recognition, object classification and time series prediction. Scaling these systems in space and time faces challenges in control complexity, size and power demand, which can be relieved by integrated optical solutions. Silicon photonics can be the disruptive technology to achieve this goal. However, the experimental demonstrations have been so far focused on spatially distributed reservoirs, where the massive use of splitters/combiners and the interconnection loss limits the number of nodes. Here, we propose and validate an all optical RC scheme based on a silicon microring (MR) and time multiplexing. The input layer is encoded in the intensity of a pump beam, which is nonlinearly transferred to the free carrier concentration in the MR and imprinted on a secondary probe. We harness the free carrier dynamics to create a chain-like reservoir topology with 50 virtual nodes. We give proof of concept demonstrations of RC by solving two nontrivial tasks: the delayed XOR and the classification of Iris flowers. This forms the basic building block from which larger hybrid spatio-temporal reservoirs with thousands of nodes can be realized with a limited set of resources.
\end{abstract}

\maketitle

\section{Introduction}
\label{sec_introduction}
The last decade has seen a flourish of new machine learning (ML) methods and applications, sometimes surpassing the human's capabilities in solving hard tasks such as object recognition \cite{russakovsky2015imagenet,buetti2019deep}, playing board games \cite{silver2016mastering} or lipreading \cite{assael2016lipnet}. Among the different approaches, Reservoir Computing (RC) has attracted growing interests due to its trade-off between performance and training complexity \cite{jaeger2001echo}. RC is a machine learning paradigm inspired by Recurrent Neural Networks (RNN) where a set of input stimuli triggers the nonlinear dynamics of a network of thousands of nodes, typically sparsely connected by fixed random weights \cite{jaeger2004harnessing}. The reservoir processes the information and maps the original input space into one of increased dimension, thus mimicking the sequence of nonlinear transformations of feed forward neural networks. In contrast with these, only the nodes in the readout layer are trained, while the network separability is guaranteed by the highly nonlinear dynamics of the reservoir. The training process is thus reduced to a linear combination of the output states of the reservoir, which can be implemented with minimal computational resources.  At the heart of RC lies the fact that loose conditions on the topology of the reservoir are required for operation \cite{maass2004computational}, which find realization in a variety of physical systems. Among these, photonic RC promises to be the ideal platform for hardware acceleration, due to an ultra-high bandwidth of operation and ease of implementation \cite{vinckier2015high,larger2017high,Appeltant2014,antonik2019large}. Excellent demonstrations exist, which  fall into two distinct categories: spatial or delayed-node reservoirs. The first has a close analogy with RNN, in which nodes are spatially distributed and can be physically interconnected, as in the network of Semiconductor Optical Amplifiers proposed in \cite{vandoorne2011parallel}, or can be software-connected, as the pixels of the spatial light modulator in \cite{bueno2018reinforcement}. Delayed node reservoirs use instead a single physical node for computation. Here, the nodes are virtual and multiplexed in time along an optical delay loop \cite{Appeltant2014}. The network connectivity is given by the inertia of the system \cite{larger2017high}, or by asynchronously sampling the readout nodes with respect to the time of injection of the inputs  \cite{paquot2012optoelectronic,vinckier2015high}. Implementations often use optoelectronic delay loops embedding nonlinear elements as Mach Zehnders or phase modulators \cite{larger2017high}, DFB lasers \cite{takano2018compact}, VCSELs \cite{vatin2019experimental} or photodetectors \cite{vinckier2015high}. Being built from off-the shelf optics and electronics, these systems operate at GHz speed and couple thousands of nodes. State of the art performances have been demonstrated on several benchmarking tests, as chaotic time series prediction, nonlinear channel equalization and speech recognition. However, to satisfy the growing demand of more challenging tasks, the number of virtual nodes and their interconnection complexity is foreseen to increase over the next years. Scaling these parameters with bulk optoelectronic solutions will soon become unpractical for both spatial and delay-loop architectures. This is why numerous RC schemes based on integrated optics have been proposed, and some of them experimentally demonstrated. Silicon photonics is the ideal platform for very large scale integration, offering the possibility to fabricate hundreds of individually addressable and reconfigurable optical elements in few cm$^2$ \cite{zevcevic20183d}. However, the number of experimental demonstrations of RC based on silicon photonics is still limited \cite{NatCom2014,Memory_OSA,takano2018compact}. Most of the proposed schemes focus on passive spatially distributed reservoirs made by small matrices of waveguides \cite{NatCom2014,katumba2019neuromorphic}, photonic crystal cavities \cite{PCrystal_Bienstman} or ring resonators \cite{denis2018all,mesaritakis2013micro,mesaritakis2015high}. The fading memory of these systems is given by the delay lines which connect the nodes, or by the photon lifetime of the cavities. This memory spans from few ps to some ns. The nonlinearity is typically provided by the use of square law photodetectors, which exploit the coherence properties of light propagating in these structures. The scale of the reservoir is limited by the loss of the delay lines and by the repeated use of splitters (coherent combining) which create the connectivity of the network \cite{katumba2018low}. The integration of single nodes with delayed feedback eliminates the interconnection problem, but also faces challenges, since lossy delay loops of the order of few ns are required to store hundreds of virtual nodes with ps separation \cite{takano2018compact,Memory_OSA}. It is natural to seek the use of hybrid spatio-temporal architectures as an optimal trade-off solution. These may adopt a small scale spatial topology with sparse connectivity, where each physical node can accommodate few hundreds of time multiplexed virtual nodes, similar to what is suggested in \cite{zhang2014integrated}. \\
\noindent In this work, we propose and experimentally validate a key element of this scheme. A RC architecture is realized with a silicon MR and time multiplexing. The input space is encoded in time bins of different light intensity, which are sequentially injected in the device to trigger a transient state in the free carrier population. The carrier dynamics defines the connections between the virtual nodes. In particular, we adopt a scheme in which the input information is nonlinearly transferred from a pump to a weak continuos wave (CW) probe laser inside the MR. This is achieved by generating carriers through Two Photon Absorption (TPA) and then by imprinting their signature on the probe intensity by Free Carrier Dispersion (FCD), which shifts the resonance frequency of the cavity. The system does not require active materials and does not rely on photodetection to induce nonlinearities, being these naturally provided by TPA. The strength of information transfer is two orders of magnitude higher than the one mediated by Kerr, and occurs at modest input power thanks to the light enhancement inside the MR. In contrast to the previously reported passive schemes \cite{NatCom2014,mesaritakis2013micro,mesaritakis2015high,PCrystal_Bienstman}, where the inputs are constrained to vary at rates above $10\,\textrm{Gbps}$ to minimize the interconnection lengths (hence the loss), the reservoir operates at the free carrier timescale, which in our case is $45\,\textrm{ns}$ ($\sim 20\,\textrm{MHz}$). This is beneficial for the driving electronics, since we can achieve an high Signal to Noise Ratio (SNR) at the readout and minimize the quantization noise on the input signal, which are both factors affecting the performance of the device \cite{soriano2013optoelectronic}. We experimentally demonstrate the RC capabilities with both binary and analog input signals. The binary task we address is the $1$-bit delayed XOR, where we achieved a minimum detectable Bit Error Rate (BER) of $1.4\times10^{-3}$ for bitrates up to $30\,\textrm{MHz}$. The analog task we tackled is the classification of the Iris dataset \cite{fisher1936use}, which yielded $(99.3\pm0.2)\%$ accuracy at a rate of $0.38\,\textrm{MHz}$. We experimentally and theoretically show that the reservoir achieves its maximum performance when the carrier and the temperature dynamics slightly mix together, which occurs at the transition edge between a stable and a self-pulsing regime \cite{johnson2006self}. Finally, we discuss how to scale the presented scheme, showing that GHz operation with thousands of nodes can be possible with limited resources. 

\section{Principle of operation}
\label{sec_principle_of_operation}
\begin{figure}[h!]
\includegraphics[scale = 0.8]{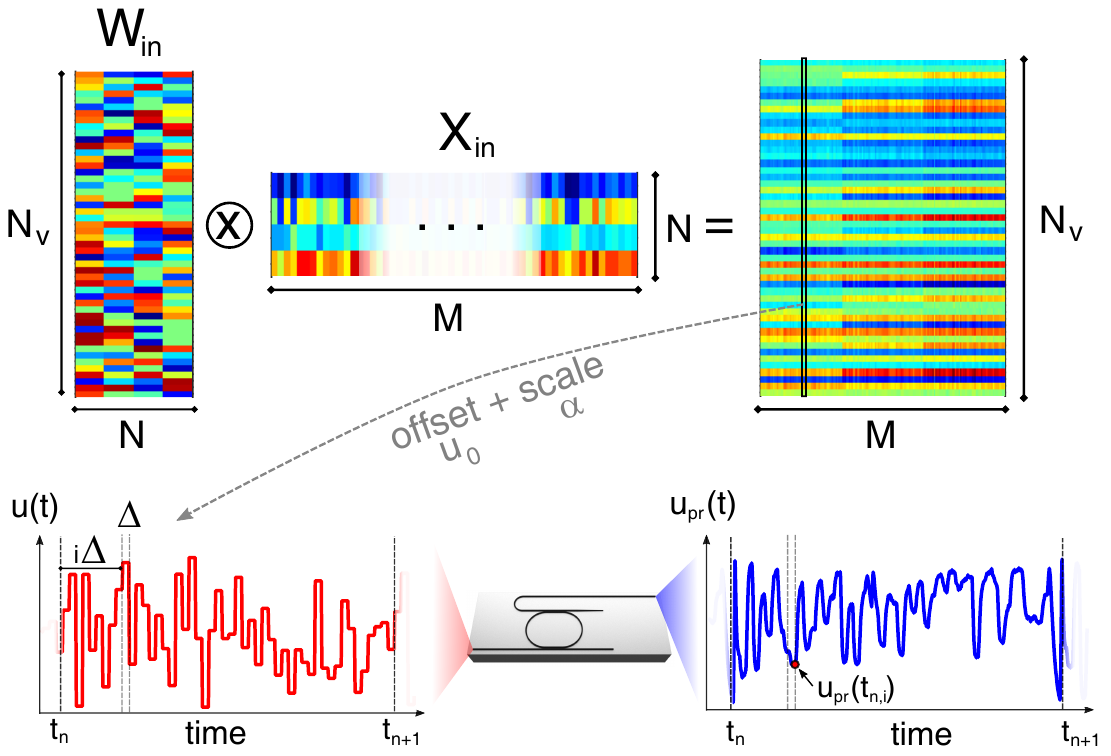}
\caption{Process flow of the encoding of the input signal. $M$ input samples $\{\mathbf{x}^{(1)},...,\mathbf{x}^{(M)}\}$ of dimension $N$ are queued on the columns of a matrix $\mathbf{X}_{\textup{in}}$. The dimension of each sample is then increased to $N_v$ using a connectivity matrix $\mathbf{W}_{\textup{in}}$. A global offset $u_0$ is applied to $\mathbf{W}_{\textup{in}}\mathbf{X}_{\textup{in}}$ to remove the negative values, and a multiplicative scale factor $\alpha$ is applied. The resulting column values represent the input pump power (red curve) $u$ of each sample, which are sequentially injected at times \mbox{$t_n = n(N_v\Delta)$} at the input port of the MR (central inset). Similarly, the values of the probe power $u_{pr}(t_{n,i})$ at times \mbox{$t_{n,i}=n(N_v\Delta)+i\Delta$}, with $i=\{1,...,N_v\}$, define the virtual nodes at the output of the Drop port of the MR. \label{Figure_0}}
\end{figure}

The architecture implements a single MR in the Add-Drop configuration as a reservoir, in which virtual nodes are defined by time multiplexing \cite{appeltant2012reservoir}. The way we construct and process the input and output layers is schematically shown in Fig.(\ref{Figure_0}).  The input signal $u(t)$, which can be a binary bit-sequence or of analog nature, is encoded in the intensity of a pump laser which is resonantly coupled to the input port of the MR. In contrast with common delayed-loop architectures \cite{Appeltant2014,paquot2012optoelectronic}, where the time of flight of light within the feedback loop sets the rate at which samples are injected and read, here there is no external feedback, so the bit duration $T=N_v\Delta$ is determined by our choice of the number of virtual nodes $N_v$ and by their time separation $\Delta$. The number $N_v$ is chosen as a trade-off between the required complexity of information processing, which generally increases with $N_v$ \cite{takano2018compact}, and the speed of operation. According to a sample-and-hold technique \cite{Appeltant2014}, each of the  $N$-dimensional input sample $\mathbf{x}^{(n)} = \{x_1^{(n)},x_2^{(n)},...,x_{N}^{(n)}\}$ at time $t_n = nT$ is kept constant during the time interval $[t_n,t_{n+1}]$, and is spread over all virtual nodes through an input connectivity matrix $\mathbf{W_{\textup{in}}}$, which defines the mask on the data. The injected pump intensity at time $t = t_{n,i}(\sigma) = nT+i\Delta+\sigma$, where $\sigma \in [0,\Delta)$ and $i=\{1,...,N_v\}$, is given by $u(t_{n,i}(\sigma)) = \alpha \sum_{k=1}^{N}W_{ik}x_k^{(n)}\theta_{n,i}(t_{n,i}(\sigma))+u_0$, in which $\theta_{n,i}(t_{n,i}(\sigma))$ is a window function of duration $\Delta$ that is equal to $1$ for $t\in [t_n+i\Delta,t_n+(i+1)\Delta)$ and $0$ elsewhere. The coefficients $\alpha$ and $u_0$ are a scale factor and an offset which respectively set the average pump power and make $u$ positive valued. In a more compact form, if we define $\mathbf{X}_{\textup{in}}=[\mathbf{x}^{(1)},...,\mathbf{x}^{(M)}]$ as the matrix of the input samples, $u(t_{n,i})$ are the entries of the matrix $\alpha(\mathbf{W}_{\textup{in}}\mathbf{X}_{\textup{in}}+u_0\mathbf{1})^T$. 
The pump wavelength $\lambda_{p}$ is set with a small detuning $\Delta \lambda_p$ with respect to one of the resonance orders $\lambda_{p,0}$ (pump resonance) of the MR. A second probe laser, at a constant power $P_{pr}\ll P_p$, is injected into the same port, and is tuned at a wavelength $\lambda_{pr} = \lambda_{pr,0}+\Delta\lambda_{pr}$, with a detuning $\Delta\lambda_{pr}$ with respect to the probe resonance wavelength $\lambda_{pr,0}$. We used $|\lambda_{pr,0}-\lambda_{p,0}| = \textup{FSR}$, where $\textup{FSR}$ is the Free Spectral Range of the MR. In the rest of this paper, we refer to $\lambda_{p,0}$($\lambda_{pr,0}$) as the "cold" pump(probe) resonance, that is the resonance wavelength of the MR in the linear operation regime, where all the nonlinear effects, which might affect the resonance frequency, are neglected. As a consequence of light confinement in the MR, the pump intensity builds up, and TPA promotes free carriers from the valence to the conduction band. This changes the refractive index of the material through FCD, inducing a rigid blue shift of all the eigenfrequencies of the MR \cite{johnson2006self}. Concurrently, the round-trip loss is increased by Free Carrier Absorption (FCA). Since $u$ varies in time, the magnitude of the FCD shift changes as well, and its dynamics is imprinted on the  probe laser intensity at the output of the MR. The latter incoherently transfers the input information from the pump to the probe beam. It is worth to note that the FCD shift caused by TPA is two orders of magnitude higher than the Kerr shift (see Appendix B), and acts as an effective intensity induced nonlinearity. The carrier lifetime $\tau_{\textup{fc}}$ is much smaller than the thermal constant $\tau_{\textup{th}}$ of the MR \cite{borghi2020OntheModeling}. Therefore, when $\Delta \sim\tau_{\textup{fc}}$, the MR temperature variations $\Delta T$ cannot follow the temporal profile of the carrier concentration. Self heating effects due to carrier relaxation are smoothed to yield a steady state value $\overline{\Delta T}$.
In this regime, the MR temperature does not participate to the reservoir dynamics, and only changes the effective detuning $\Delta \lambda_p$($\Delta \lambda_{pr}$) of the pump(probe) resonance. However, as discussed in Section \ref{sec_experimental_realization}.\ref{sec_badone}, this holds only at low input power. When the average power is raised above a certain threshold, the MR starts to self-pulse \cite{johnson2006self}, and the temperature dynamics impacts the processing capabilities of the nodes. 
\begin{figure*}[t!]
\centering\includegraphics[scale = 0.65]{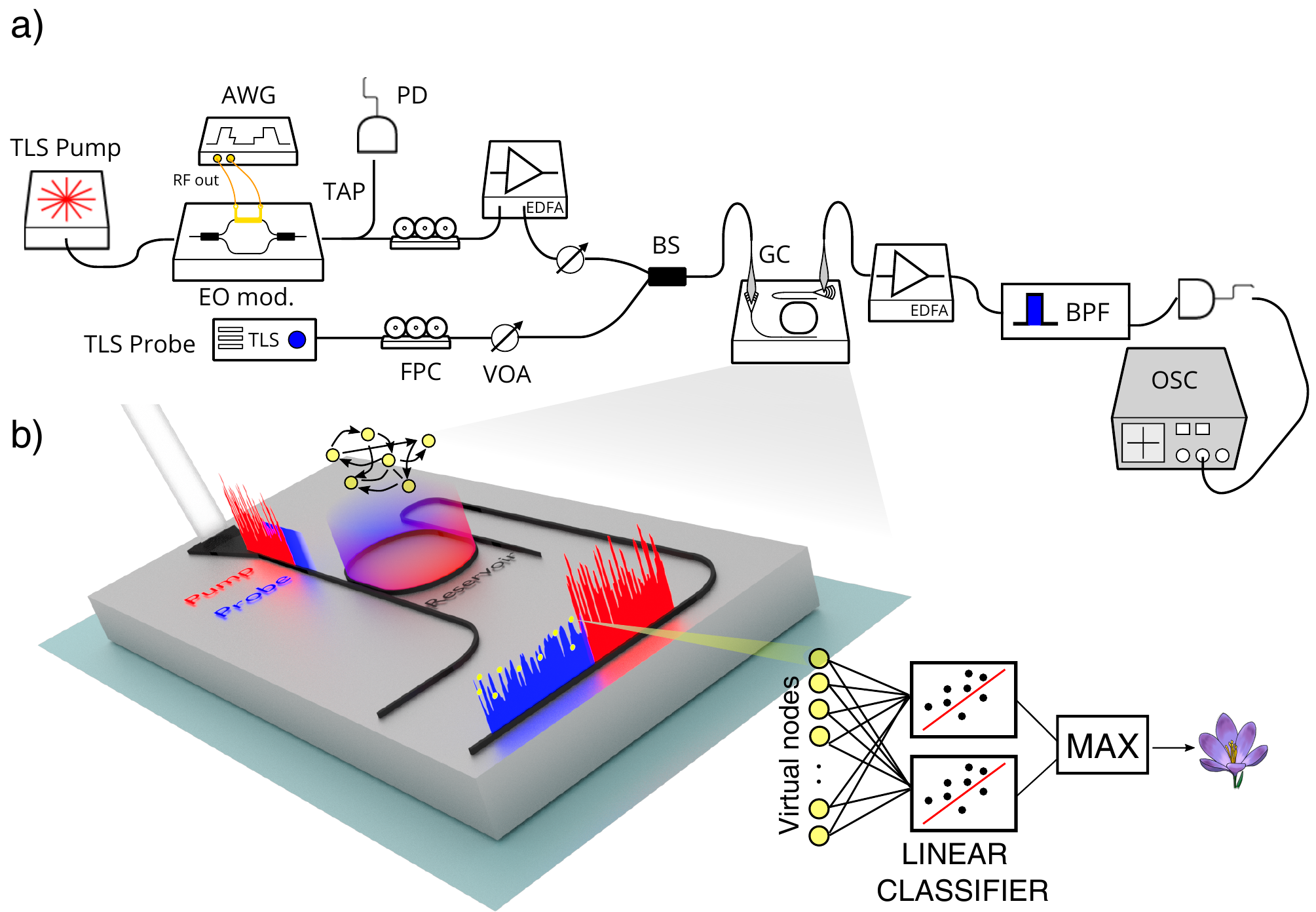}
\caption{(a) Sketch of the experimental setup. TLS = Tunable Laser Source, AWG = Arbitrary Waveform Generator, EO mod. = Electro Optic modulator, FPC = Fiber Polarization Controller, VOA = Variable Optical Attenuator, PD = Photodiode, EDFA = Erbium Doped Optical Amplifier, BS = Beam Splitter, GC = Grating Coupler, BPF = BandPass Filter, OSC = Oscilloscope. b) Device layout and logical information flow. The intensity modulated pump (red) and the CW probe (blue) are injected into the input GC. The incoherent transfer of information from the pump to the probe occurs within the resonator (reservoir), where the different virtual nodes (yellow dots) interact and process the input data. The probe exits from the Drop port, carrying the result of the  computation. Virtual nodes are sampled and sent into several linear classifiers, each trained to recognize a specific class. The decision making process is based on a winner takes all scheme. \label{Figure_1}}
\end{figure*}
\noindent We can get an insight on the internal structure of the reservoir by assuming small perturbations of the input power $u(t)$ with respect to a reference value $\overline{u(t)}$. Here,  small means that the corresponding free carrier change $\Delta N-\overline{\Delta N}$ around the reference value $\overline{\Delta N}$ induces variations of the transmitted pump/probe intensity which are linear with $\Delta N$. As derived in Appendix A, within this approximation the probe power $u_{pr}$ at the Drop port of the resonator is given by:
\begin{equation}
\label{eq:1}
\begin{split}
u_{pr}(t) & = c_0+c_1\int_{-\infty}^{t}e^{-\left (\frac{t-\xi}{\tau_{\textup{fc}}}\right )}u^2(\xi) d\xi +  \\
{} & + c_2 \int_{-\infty}^{t}e^{-\left (\frac{t-\xi}{\tau_{\textup{fc}}}\right )}u^2(\xi)u_{pr}(\xi)d\xi,
\end{split}
\end{equation}
where the definitions of $c_0$, $c_1$  and $c_2$ are given in Appendix A. The first contribution to the virtual node state at time $t$ is quadratic in the pump power $u$, and depends on the past inputs with weights that exponentially decrease with time. This short term memory is of the order of $\sim3\,\tau_{\textup{fc}}$, i.e., the time required by free carriers to reach a steady state after an abrupt change of the pump power. The memory of this system is thus intrinsically nonlinear. This is because the input information $\mathbf{x}^{(n)}$ is instantaneously (fs scale) transferred to carriers through nonlinear absorption, temporarily processed from a slower (ns) dynamics, and subsequently imprinted to the output probe. Similarly to what is done in \cite{appeltant2012reservoir}, one can approximate the integral in the first term with a discrete sum, and find the recursive virtual node relation $u_{pr}(t_{n,i}) = \Delta c_1\sum_{k=0}^{m}\left( \eta^k u(t_{n,i-k})^2 \right)+\eta^m u_{pr}(t_{n,i-m})$ where $\eta = e^{-\left ( \frac{\Delta}{\tau_{\textup{fc}}}\right )}$ (for simplicity, the second term in Eq.(\ref{eq:1}) has been neglected, but this does not alter our conclusions). This highlights the type of connectivity between the virtual nodes of the reservoir, which are organized in a chain-like topology: node $u_{pr}(t_{n,i})$ is coupled with its adjacent $u_{pr}(t_{n,i-1})$ with a coupling strength $\eta$. The second term in Eq.(\ref{eq:1}) is a nonlinear coupling between virtual nodes. As shown in Appendix A, this term arises from the the resonance shift induced by FCD. This shift modifies the pump power circulating in the MR, which in turn affects the carrier generation rate. As a result, a recursive relation between these quantities is settled, for which the second term in Eq.(\ref{eq:1}) accounts for the leading term. When the change in $u_{pr}$ is no more linear with $\Delta N-\overline{\Delta N}$ due to large FCD induced resonance shifts, Eq.(\ref{eq:1}) loses validity and the virtual node interaction becomes increasingly complex. The general structure of the network is however maintained, with TPA and inter-node coupling providing nonlinearity, while carrier recombination a short term memory to the reservoir. Virtual nodes $u_{pr}(t_{n,i})$ are sampled at the Drop port, and arranged into a state matrix $\mathbf{X} = [\mathbf{u}_{pr}^{(1)},...,\mathbf{u}_{pr}^{(M)}]^{T}$, where $\mathbf{u}_{pr}^{(k)}$ is the column vector collecting the $N_v$ virtual nodes corresponding to the $k^{\textup{th}}$ input sample $\mathbf{x}^{(k)}$. The action of the reservoir is to project the initial set of predictors $\mathbf{x}^{(k)}$ with corresponding observables $\mathbf{y}^{(k)}$ (in general, a Q-dimensional vector) to a higher dimensional space $\mathbf{x}^{(k)}\rightarrow \mathbf{u}_{pr}^{(k)}$ where they ideally should be linearly separable. Then, the aim is to find a $(N_v\times Q)$ weight matrix $\mathbf{W_{\textup{out}}}$ which realizes $\mathbf{Y}=\mathbf{X}\mathbf{W_{\textup{out}}}$, in which the observables are  arranged in the  $M\times Q$ matrix $\mathbf{Y}$. We accomplish this task by regularized least squares (Ridge regression), where the regularization parameter $\lambda$ is determined by a $5$-fold cross validation \cite{tikhonov1977solutions}. The output of this procedure is a matrix $\mathbf{\Tilde{W}_{\textup{out}}}$ which minimizes the regularized least square error $\sum_{k=1}^{M}||(\mathbf{Y}-\mathbf{\Tilde{W}_{\textup{out}}}\mathbf{X}) ||^2+\lambda^2||\mathbf{\Tilde{W_{\textup{out}}}}||^2$. In case of supervised learning, as we deal in this paper, $\mathbf{y}^{(k)}$ is a $Q$-dimensional binary variable, with $1$s identifying the assignment category of $\mathbf{x}^{(k)}$ and $0$s elsewhere. Given $\mathbf{\Tilde{y}}^{(k)}$ and $\mathbf{y}^{k}$, the way decisions are handled depends on the task, and will be treated separately in Sections \ref{sec_xor} and \ref{sec_iris_species}.
\section{Experimental realization}
Our experimental apparatus is sketched in Fig.\ref{Figure_1}(a). The pump is a C-band, CW tunable laser (Pure Photonics) which is intensity modulated by an electro-optic IQ modulator (IxBlue MXIQ-LN-$30$) to realize the desired input pump waveform $u(t)$. The modulator is driven by a $65\,\textrm{Gs}$ Arbitrary Waveform Generator (AWG) from Keysight, whose output is amplified by a high bandwidth amplification stage (IxBlue DR-AN-$28$-MO), providing the necessary voltage swing of $V_{\pi}\sim7\,\textrm{V}$ to exploit the full dynamic range of the modulator. In order to claim that the nonlinear transformation on the input samples is solely coming from the MR, we did a pre-compensation of the signal out of the AWG, which eliminates the $\cos^2$ dependence of the modulator intensity response with the applied voltage. A tap of $10\%$ is placed at the output of the modulator to monitor the input pump, which is detected by a fast photodiode (Thorlabs DXM20AF). The pump is amplified to a fixed level of $20\,\textrm{dBm}$ by an Erbium Doped Optical Amplifier (EDFA, IPG Photonics) and the power regulated by an electronic Variable Optical Attenuation (VIAVI mVOA-C1). A second CW probe laser (Pure Photonics) is combined with the pump before being injected into the input port of the MR using a single mode fiber and a Grating Coupler  ($\sim\,3.5\,\textrm{dB}$ loss). The MR under test has a racestrack shape, with a radius of $7\,\mu\textrm{m}$ and a waveguide cross section of $450\,\textrm{nm}\times220\,\textrm{nm}$. Two bus waveguides are coupled to the straight sections of the MR,  with a gap of $250\,\textrm{nm}$ and a coupling length of  $3\,\mu\textrm{m}$. The device is described in great details in \cite{borghi2020OntheModeling}. From the low-power transmission spectra recorded at the Through port, we extracted an intrinsic quality factor (Q) of $Q_{\textup{i}} = 1.11(8)\times10^5$ and a loaded Q of $Q_{\textup{L}} = 6.5(2)\times10^{3}$. We use two adjacent resonance orders to resonantly couple the pump and probe laser, which have respectively a cold resonance wavelength of $\sim1549\,\textrm{nm}$ and $\sim1538\,\textrm{nm}$. 
Light at the output of the Drop port is amplified by a second EDFA (Thorlabs), and a tunable band-pass filter ($0.8\,\textrm{nm}$ bandwidth) removes the spontaneous emission and directs the probe or the pump laser to the output photodiode (Thorlabs D400FC, $1\,\textrm{GHz}$ bandwidth). A $4\times40\,\textrm{Gs}$ oscilloscope (LeCroy SDA 816Zi-A) records the input pump waveform $u(t)$ and the output probe $u_{pr}(t)$. Virtual nodes are sampled from $u_{pr}(t)$ at times $t_{n,i}$ and queued to form the state matrix $\mathbf{X}$. Training and validation are performed off-line using Matlab. 
\label{sec_experimental_realization}
\subsection{Binary input: 1-bit delayed XOR}
\label{sec_xor}
The first task we tested is the 1-bit delayed XOR \cite{NatCom2014}. Given the virtual node vector $\mathbf{u}_{pr}^{(k)}$ at time $t_k$, the goal is to predict the result of the XOR operation between the bits $x^{(k)}$ and $x^{(k-1)}$. 
\begin{figure}[h!]
\centering \includegraphics[scale = 0.72]{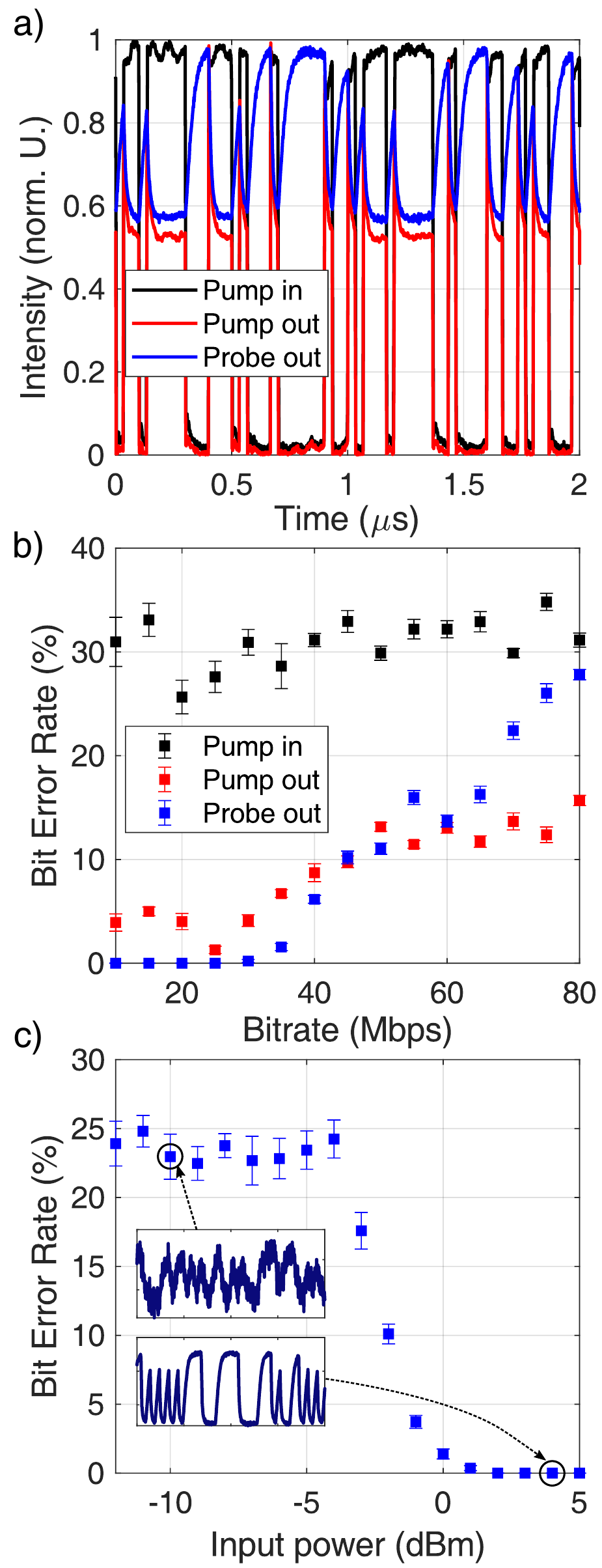}
\caption{(a) Examples of waveforms processed during the XOR task. The pump laser driving the input port of the MR is shown in black, while the pump and probe outputs from the Drop port are respectively shown in red and blue. (b) Bit Error Rate (BER) as a function of the bitrate for the 1-bit delayed XOR task. Black dots use a predictor matrix $\mathbf{X}$ whose entries are sampled from the input pump power. Red and blue dots use respectively predictors sampled from the pump and the probe traces at the Drop port of the MR. In all the three cases, the average pump power is set to $3\,\textrm{dBm}$. (c) Bit Error Rate as a function of the average pump power for a fixed bitrate of $20\,\textrm{Mbps}$. The insets show details of the probe waveform at the pump powers $-10\,\textrm{dBm}$ and $4\,\textrm{dBm}$. \label{Figure_2}}
\end{figure}
The input variable can assume two values, $1$ (full transmission of the modulator) and $0$ (full extinction). In this task, we use $N_v=3$ virtual nodes, and the connectivity mask is simply a $3\times 1$ column vector of $1$s. This implies that no mask is applied to the input data, so that $u(t_{n,i}(\sigma))=x^{(n)}\theta(t_{n,i}(\sigma))$. Despite the apparent simplicity, this task unveils most of the features of the reservoir architecture: the nonlinear transformation of the inputs, the presence of a fading memory, and the ability to deal with binary inputs. In this experiment, we fix the average power of the pump at the input to $3\,\textrm{dBm}$, which is sufficient to trigger TPA and carrier generation, and its detuning to $\Delta\lambda_p= 60\,\textrm{pm}$. The pump intensity is  modulated with $\sim 18\,\textrm{dB}$ of extinction, and the input is a Pseudo Random Binary Sequence (PRBS) of bitrate $B$, which we varied from $10\,\textrm{Mbps}$ to $80\,\textrm{Mbps}$. The probe power is set to $P_{pr} = -3\,\textrm{dBm}$, which is sufficiently low to not alter the resonance shift imparted by the pump, and its detuning to $\Delta\lambda_{pr} = 60\,\textrm{pm}$. An example of waveforms at the input and output of the MR for a bitrate of $30\,\textrm{Mbps}$ are shown in Fig.\ref{Figure_2}(a). Whenever the input pump changes from a low to an high state, a transient exponential decay is observed at the Drop port, which is due to the blue shift of the resonance as a consequence of carrier generation and accumulation. The cold cavity detuning $\Delta\lambda_{p(pr)}$ is increased by FCD, determining a decrease of the transmittance, which also occurs for the probe output. The opposite trend is observed, in the probe trace, during the transition from a high to a low state. Here, carrier recombines and the detuning $\Delta\lambda_p(pr)$ decreases, pulling the resonance back to the cold position. All the transient phenomena occur at the characteristic timescale of $\tau_{\textup{fc}}=45\,\textrm{ns}$, i.e., the value of the carrier lifetime we measured in one of our previous works \cite{borghi2020OntheModeling}.\\
\noindent Virtual nodes are obtained by sampling these traces at $10\,\textrm{Gbps}$, smoothing them with a moving average filter ($10$ points), and downsampling them at times $\Delta = (N_v B)^{-1}$. After smoothing, the Signal to Noise ratio (SNR) increases from $18\,\textrm{dB}$ (from the raw data acquisition) to $\sim20\,\textrm{dB}$. Figure \ref{Figure_2}(b) shows the Bit Error Rate (BER) as a function of the bitrate for different predictor matrices $\mathbf{X}$, whose entries are respectively populated by sampling the virtual node values from the input pump intensity (black), the output pump power (red) and the output probe power (blue). To extract the BER, $6500$ bits are used for training and $3500$ for validation. Errorbars are obtained by partitioning the validation set into $5$ non-overlapping batches, over which the BER is evaluated. The error is estimated as the standard deviation over the different batches. The output of the linear classifier $\Tilde{\mathbf{Y}}$ is digitized by applying a threshold, which optimizes the BER at each bitrate. The XOR task is never solved by sampling virtual nodes from the bare input pump. The average BER is $\sim30\%$ for all the bitrates. This agrees with the fact that the task is not linearly separable and requires at least one bit of memory to be performed. 
Virtual nodes sampled from the pump at the output of the Drop port improve the BER, which is however still bounded above $\sim 5\%$. The reason behind this is that bits which are zero do not carry optical power, so these virtual nodes are sampled from the background noise. This issue is solved by using the probe signal. Indeed, error free operation is achieved for bitrates lower than $25\,\textrm{Mbps}$. Since we use $700$ bits for each partition, error free means that the actual BER is lower than $1.4\times10^{-3}$. At higher rates, carriers are no more able to react sufficiently fast to follow the pump power variations, so the BER increases.  The ability to solve the delayed XOR demonstrates the presence of memory and nonlinearity in the reservoir. This also illustrates that the use of a probe beam is beneficial for tasks where the input is binary and with full modulation. It helps to improve the separability of the reservoir \cite{mesaritakis2013micro}. To investigate the influence of the input power on the BER, we fixed the bitrate to $20\,\textrm{Mbps}$, and we changed the average input power of the pump $P_p$. The calculated BER is shown in Fig.\ref{Figure_2}(c). The error rate is almost  power-insensitive and equal to $\sim25\%$ for $P_p<-4\,\textrm{dBm}$, while it drops below $1\%$ at $P_p=0\,\textrm{dBm}$. At $P_p>2\,\textrm{dBm}$, error-free operation is achieved. The low and high power insets shown in Fig.\ref{Figure_2}(c) explain why the BER improves by increasing power. At $P_p=-10\,\textrm{dBm}$, few carriers are generated, so the probe is weakly perturbed from its original CW state and virtual nodes acquire a noisy flat distribution of values. At $P_p=4\,\textrm{dBm}$, the FCD shift is source of more variability in the virtual nodes states, improving the input separability of the reservoir. We also averaged several traces at low power in order to compare the performance at the same SNR of the ones at high power, but we only observed a minor improvement of the BER. This confirms that the poor performance measured at low power is due to the reduced virtual node variability. 

\subsection{Analog input: Iris species recognition}
\label{sec_iris_species}
\begin{figure*}[t]
\centering\includegraphics[scale = 0.8]{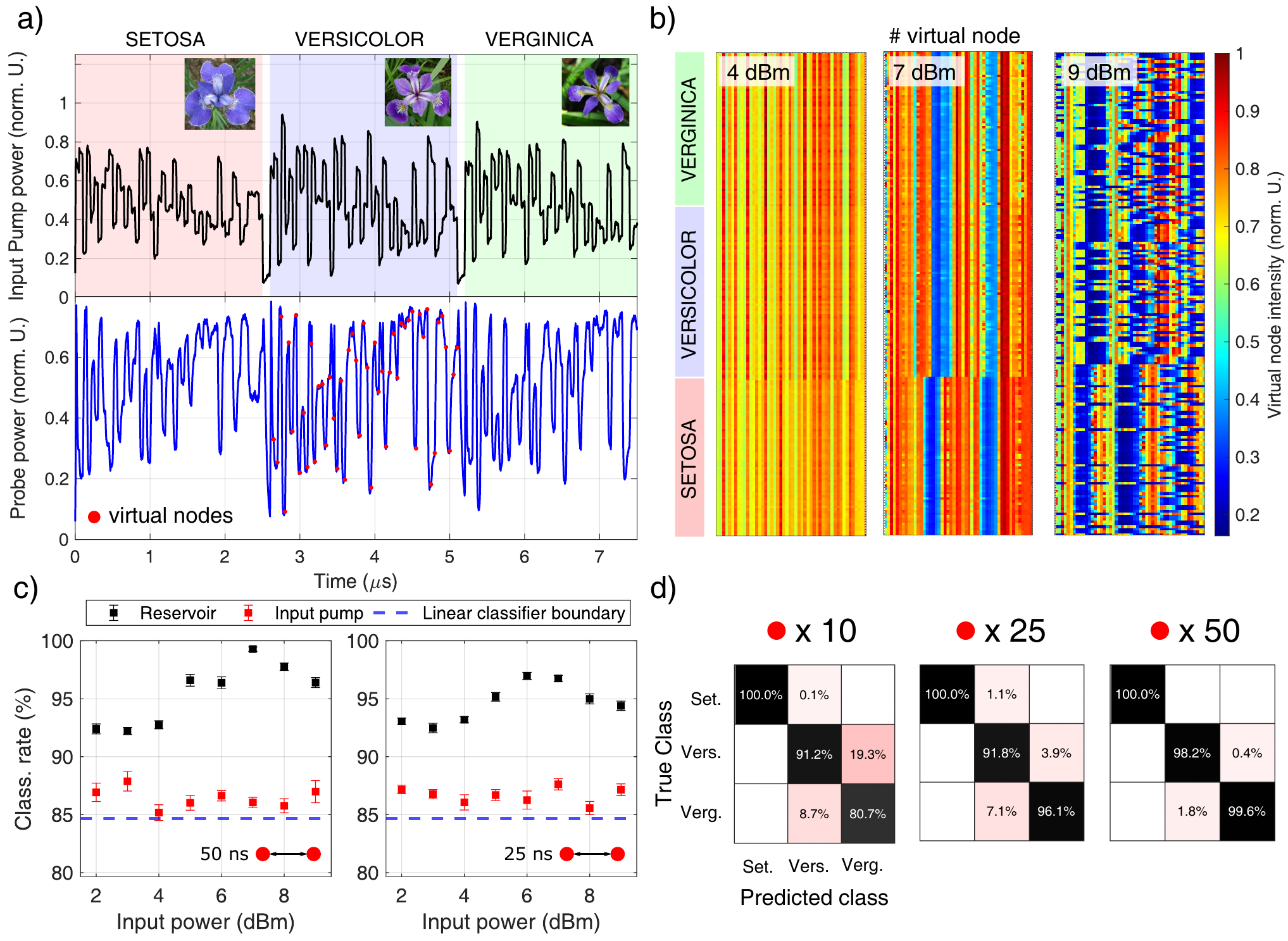}
\caption{(a) Examples of waveforms of the input pump (top panel, in black) and of the probe at the output of the MR (bottom panel, in blue) for $N_v = 50$ and a bitrate of $20\,\textrm{Mbps}$. Different background colours highlight the subspecies of the flower. A representative picture of them is shown in the insets of the top panel. Virtual nodes sampled from the probe trace are indicated with red dots on the central band. (b) Maps of the (normalized) intensity of the different virtual nodes ($N_v = 50$, bitrate $20\,\textrm{Mbps}$) sampled from the output probe, acquired at the pump powers of $4\,\textrm{dBm}$ (left), $7\,\textrm{dBm}$ (center), and $9\,\textrm{dBm}$ (right). These are $189$ flower samples vertically stacked in each map. Even if they are randomly injected at the input port during the training and test phase, they are shown grouped together in the three subspecies, as indicated by the labels on the left. This highlights the distinction between the classes.  (c) Classification rate as a function of the average input power of the pump for $N_v = 50$ and bitrates of $20\,\textrm{Mbps}$ (left) and $40\,\textrm{Mbps}$ (right). Black scatters use virtual nodes sampled from the output probe while red scatters from the input pump. The dashed blue line is the lower bound in the classification rate obtained by feeding the input samples into a linear classifier, without  electro-optic conversion. (d) Confusion charts for a MR reservoir which uses $10$ (left), $25$ (center), and $50$ (right) virtual nodes. The bitrate is $20\,\textrm{Mbps}$. \label{Figure_3}}
\end{figure*}
The second task we tested is the classification of the Iris flowers, a well known dataset \cite{fisher1936use} in the field of machine learning. Here, the goal is to classify a given Iris flower into one of the three possible subspecies: \emph{Setosa}, \emph{Versicolor} and \emph{Verginica}. The classification is based on four real inputs $\textbf{x}$, physically corresponding to the length and width of the petals and sepals of the flower. All the four inputs are required to solve the task, since the flowers can not be linearly classified on the basis of a single feature. Moreover, even by using all the four inputs, only one species (\emph{Setosa}) is linearly separable from the others, making the task nontrivial. As described in Section \ref{sec_principle_of_operation}, the dimension of the original input space is increased from $4$ to $N_v$ by implementing the transformation $\mathbf{x}\rightarrow \mathbf{W}_{\textup{in}}\mathbf{x}$, where the connectivity matrix $\mathbf{W}_{\textup{in}}$ has dimension $N_v\times 4$. An offset is  applied to the transformed input to remove the negative values, and the result is scaled to reach the desired target average power $P_p$. The sequence is then encoded in the pump intensity, and sequentially injected to the input port of the MR. We tested different numbers of virtual nodes $N_v = \{10,25,50\}$, and two values of time separation $\Delta = \{25,50\}\,\textrm{ns}$. Each flower sample is sequentially encoded in the pump intensity  $u(t)$ and injected into the input port of the MR. Different flowers are separated by a delay of $\delta = 100\,\textrm{ns}$, during which the pump power is turned off to quench the carrier dynamics. This avoids undesired crosstalk of information between neighbouring flowers. The original dataset contains $150$ flowers, with an equal number of representatives of each class. Each instance is injected into the chip is randomly extracted from the dataset. Figure \ref{Figure_3}(a)(top panel) shows an example of pump waveform at the input of the MR for three flower instances of different subspecies. Each is made by $50$ steps, which corresponds to the virtual nodes. The low speed of operation allows to use the full resolution of the digital to analog converter of the AWG ($8$ bit), imprinting on the optical power trace the fine details of each flower sample. The associated probe waveform, after being processed by the MR reservoir, is shown in Fig.\ref{Figure_3}(a)(bottom panel). With respect to the XOR task, where the input is binary, the multilevel excitation imparts a richer dynamics on the output probe, which results in an enhanced variability between the virtual nodes. The reservoir response to different flower samples is represented in the maps of Fig.\ref{Figure_3}(b), which shows the intensities of the $50$ virtual nodes for $189$ flower instances (vertically stacked). In Fig.\ref{Figure_3}(c) we report the classification rate for $20\,\textrm{Mbps}$ and $40\,\textrm{Mbps}$ as a function of the average input power. Categories are assigned by training multiple linear classifiers, one for each subspecies, and decisions are made on the basis of a winner takes all scheme \cite{larger2017high}.  The best performance is obtained for $20\,\textrm{Mbps}$ and a pump power of $P_p = 7\,\textrm{dBm}$, yelding a classification rate of $(99.3\pm0.2)\%$. In comparison, the classification rate obtained by sampling the virtual nodes from the input pump is below $87\%$. This value is slightly above the theoretical limit of $84.66\%$, obtained by directly feeding the input predictor matrix $\mathbf{W}_{\textup{in}}\mathbf{X}_{\textup{in}}$ into a linear classifier. This is probably due to some residual nonlinear distortion in the electro-optic conversion. The rate of classification obtained by a single MR exceeds the accuracy achieved by a more complex integrated system (three layered feed forward neural network) presented in a recent work \cite{IRIS_Integrato}, and has competitive performance with software based classification algorithms \cite{pinto2018iris}. 
The classification rate in Fig.\ref{Figure_3}(c) increases from low to high power, is maximum at $7\,\textrm{dBm}$ and then decreases for higher powers. We can understand this trend by observing the virtual node intensity in the maps of Fig.\ref{Figure_3}(b). 
At low powers ($P_p = 4\,\textrm{dBm}$ in Fig.\ref{Figure_3}(b)), the MR reservoir shows poor separability of the inputs, as witnessed by the quite flat virtual node intensity distribution observed both within the same flower (horizontal lines) and between different subspecies (vertical lines). As discussed for the XOR task, this is related to the weak perturbations imprinted on the CW probe from the pump intensity at low powers. At the highest classification rate of $P_p = 7\,\textrm{dBm}$ (Fig.\ref{Figure_3}(b)), the virtual node distribution is richer, with a relative variation that can exceed $70\%$ (see Fig.\ref{Figure_3}(b) at $P_p = 7\,\textrm{dBm}$). We also observe the creation of two bands, where the virtual node intensity is around the $35\%$ of its maximum value. As discussed in Section \ref{sec_badone},  these form as a consequence of the interplay between thermal and free carrier dispersion. At high power ($P_p = 9\,\textrm{dBm}$ in Fig.\ref{Figure_3}(b)), multiple bands of low transmittivity are formed, which lack synchronization with the rate of flower injection. As discussed in Section \ref{sec_badone}, within the bands the MR is off resonance, hence the information can not be efficiently imprinted from the pump to the probe, effectively "burning" these virtual nodes from the reservoir. This explains the drop in the classification accuracy at high powers.\\
\noindent The net flower classification rate can be calculated as $(N_v\Delta+\delta)^{-1}$, which for $N_v=50$ and $\Delta = 50\,\textrm{ns}$, i.e., the configuration showing the best accuracy, is $0.38\,\textrm{MHz}$. We tried to scale down to $N_v=25$ and $N_v = 10$ in order to improve the speed of operation, but as shown by the confusion charts in Fig.\ref{Figure_3}(d), the classification accuracy drops to $(93.6\pm0.3)\%$ for $N_v = 25$ and $(89.2\pm0.2)\%$) for $N_v = 10$. We did not observe any significant improvement by scaling up to $N_v=100$.  

\subsection{Enhancing the reservoir performance by mixing thermal and free carrier dynamics}
\label{sec_badone}
In order to explain the behavior of the classification rate as a function of the input pump power in Fig.\ref{Figure_3}(c), we performed numerical simulations. They are based on the theoretical model of \cite{borghi2020OntheModeling}. Specifically, we computed the time evolution of the complex field of the MR, including TPA, FCA, FCD and the thermo optic dispersion (TOE).  Whenever possibile, the parameters are taken from the experiment. A more comprehensive description of the model can be found in \cite{borghi2020OntheModeling}. We fix the number of virtual nodes to $50$, the bit rate to $20\, \textrm{MHz}$ and the quenching time to $\delta=100\,\textrm{ns}$. 
\begin{figure}[t!]
\centering\includegraphics[scale =0.53]{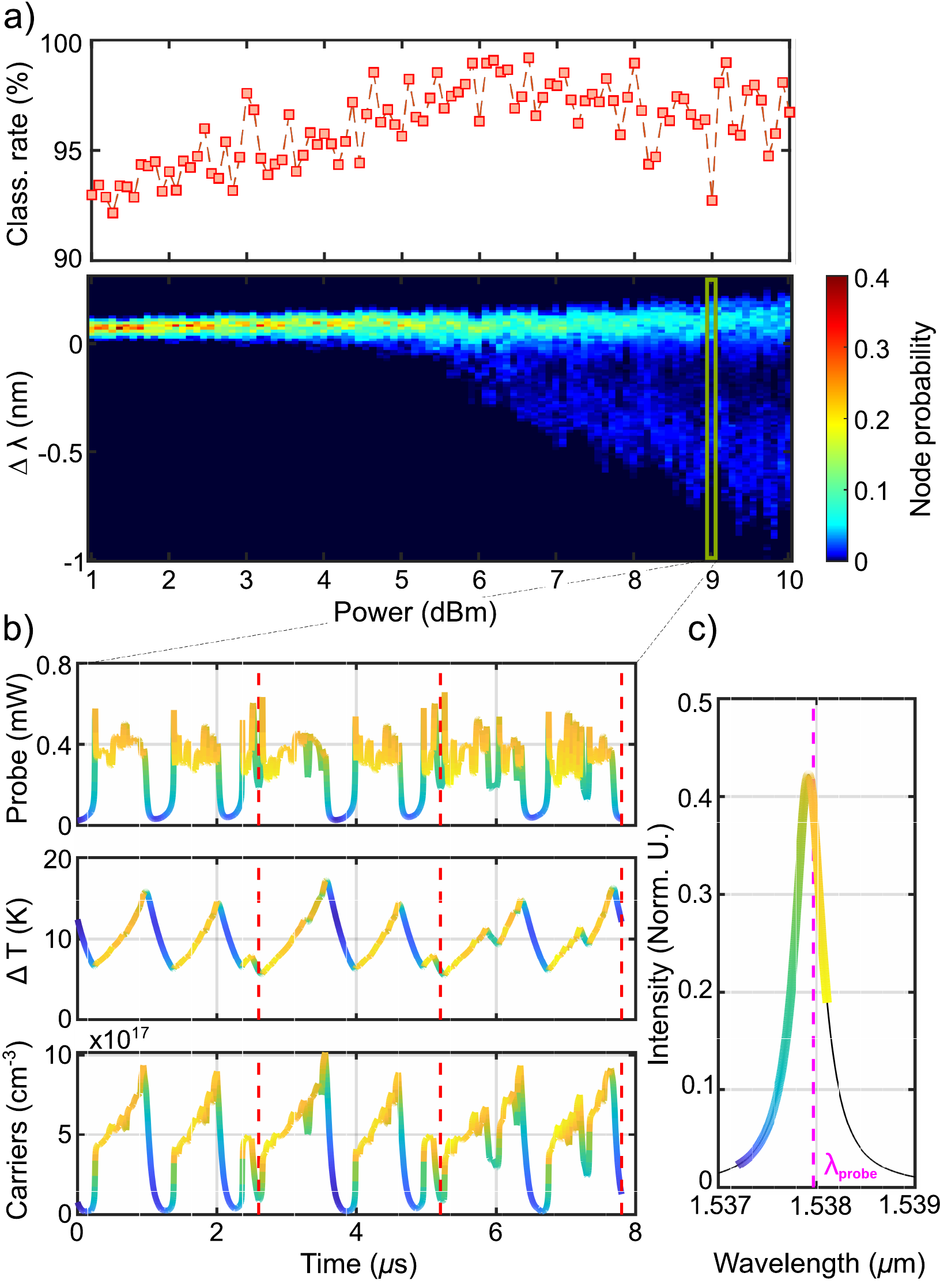}
\caption{(a) Top: numerical simulation of the Iris dataset classification rate as a function of the input power. Bottom: probability distribution of the virtual node detuning $\Delta\lambda$ as function of the input pump power. (b) Simulated  probe power at the Drop port of the MR (top), temperature variation (middle) and free carrier concentration for $P_p=9\,\textrm{dBm}$. The red dashed lines mark the injection times of three different flower samples. The color code indicates the instantaneous detuning of the probe, whose relation with the Drop intensity is shown in panel (c). (c) Normalized intensity at the Drop port of the MR as a function of wavelength. The dashed line marks the wavelength of the probe laser. The color code is used to label the instantaneous detuning of the time traces in panel (b). \label{Figure_4}}
\end{figure}
\noindent The top panel in Fig.\ref{Figure_4}(a) shows the simulated classification rate as a function of the input power,  estimated from the whole Iris dataset ($150$ flowers).  We add white noise to the simulated output of the MR to set the SNR to $18\,\textrm{dB}$, which is comparable to the one of the experiment. Each point is the result of an average over $1000$ realizations. We notice a very good agreement with the experimental data in Fig. \ref{Figure_3}(c), with a maximum of classification occurring around $7\textrm{dBm}$. We investigate the origin of this trend starting from the color map reported in the bottom panel of Fig. \ref{Figure_4}(a). Here, for each input power, we calculate the probability that a virtual node is sampled at the probe resonance detuning $\Delta\lambda$ from the cold cavity condition. At low power,  all virtual nodes are sampled around the cold resonance detuning, which in our case lies at $\Delta\lambda=60\,\textrm{pm}$. As we raise the input power up to about $5\, \textrm{dBm}$, FCD shifts the resonance frequency towards the blue, further increasing the initial detuning. Starting from $6\,\textrm{dBm}$, we observe that the variance increases, and that there is a small probability to sample virtual nodes at negative detunings. For  $P_p>6\, \textrm{dBm}$, the distribution is bimodal, i.e, nodes are concentrated around two bands: a narrow one at positive $\Delta\lambda$ and a broad one at negative $\Delta\lambda$. This is a signature that the temporal evolution is not only related to free carriers, where the temperature acts as a constant background. In fact, at these powers, we are entering a regime where carriers and temperature dynamics mix together. This is shown in the panels of Fig.\ref{Figure_4}(b), which refer to $P_p=9\,\textrm{dBm}$. Here we report, as a function of time, the output power of the probe (top), the temperature difference between the MR and its surroundings (middle), and the free carrier concentration (bottom) for three consecutive flower samples (the injection times are marked with red dashed lines). The color code indicates the instantaneous detuning of the probe, whose relation with the Drop intensity is shown in Fig.\ref{Figure_4}(c). 
When a flower sample is injected, the temperature of the MR increases, but at low powers TOE is counteracted by FCD. At high power TOE prevails,  and the wavelength detuning of the cavity can eventually become negative. When this occurs, a positive feedback is established between the decrease of the carrier concentration and the decrease of MR energy. This corresponds to the transition from the high power stability branch of the carrier bistability curve to the one at low power \cite{zhang2013multibistability}. After the transition, the MR lies off resonance \cite{johnson2006self}. The detuning is now negative due to the residual thermo optic shift. This originates the bands of low transmittivity in Fig.\ref{Figure_3}(b) and in Fig.\ref{Figure_4}(b). Virtual nodes sampled within these bands form the second broad peak in the probability distribution in Fig.\ref{Figure_4}(a). As the MR cools down, the resonance moves towards the pump(probe) wavelength, gradually increasing the probe signal. A transition from the low power stability branch of the carrier bistability curve to the one at high power restores the initial conditions, so that virtual nodes are no more sampled within the band. This interplay of thermal and free carrier effects would generate cavity self-pulsing in case of CW input powers \cite{johnson2006self}. It turns out that, for a limited range of powers centered around $7\,\textrm{dBm}$, the presence of the bands increases the classification rate. This finds explanation in the enhanced variability that the virtual nodes values assume. Indeed, in this regime, they efficiently sample all the probe resonance. Moreover, quenching the MR for $\delta=100\,\textrm{ns}$ allows to reset the initial conditions for the temperature and the carrier population between  consecutive flower samples, creating bands always in the same time slot (see Fig.\ref{Figure_3}(b), $P_p=7\,\textrm{dBm})$.
If the power is increased above $7\,\textrm{dBm}$, bands get wider in time and many of them form within the same flower sample. In these conditions, a large number of virtual nodes are sampled within bands, where the cavity lies off resonance and hence no information can be transferred from the pump to the probe beam. Furthermore, the quenching time is no more sufficient to reset the initial conditions, so as bands form at random times, turning the probe output to chaotic (see Fig.\ref{Figure_3}(b), $P_p=9\,\textrm{dBm}$)  \cite{mancinelli2014chaotic}. 


\section{Discussion}
\label{sec_discussion}
In the above sections, we demonstrated the use of a single integrated MR as a reservoir, in which the core computing power is provided by the complex coupled dynamics of free carriers, TPA and temperature. 
The architecture is simple, and as such faces limitations which can be however mitigated by different engineering solutions. While a data processing speed of few $\textrm{MHz}$ could be of interest for some specific tasks as speech recognition \cite{Phoneme_Rec}, event detection or remote control \cite{antonelo2008event}, it is not as competitive as electronic solutions. In fact, the optical implementation of machine learning is not only of interest for the low power consumption and minimal latency, but especially  for the large amount of information that can be processed, at rates in excess of Gbps \cite{PNN_Survey}. We envisage that there are at least three ways to raise the performances of our scheme. First, the MR can be used to form more advanced spatial topologies, i.e. matrices/arrays of MR. As an example, using resonators with a moderately high Q factor of $6.5\times10^4$, a network made by $10$ MR can accomodate $500$ virtual nodes and operates with $5\,\textrm{mW}$ of power. Moreover, small sequences of coupled resonators have shown very complex dynamics under thermal and free carrier nonlinearities \cite{mancinelli2014chaotic}. In principle, this suggests that the same performance could be obtained with less virtual nodes, hence increasing the speed of operation. Additionally, one could multiplex several spatial inputs \cite{katumba2017multiple} and coherently combine them on chip using linear optical elements, such as arrays of Mach-Zehnder interferometers arranged in a universal scheme \cite{DeepNature}. This hybrid spatio-temporal architecture combines a simple hardware implementation, typical of delayed nodes RC, with the compactness of a silicon photonic chip.  Second, to increase the speed of operation, the intrinsic time response of the RC can be improved by reducing the carrier lifetime. Indeed, our devices have unusually long recombination times compared to what is reported in literature \cite{xu2006carrier}. A control of the lifetime can be obtained by embedding carrier depletion modulators within the MR \cite{Electrooptical}. Third, by exploiting different resonance orders, we can implement  wavelength-division multiplexing \cite{xu_print,WavDivMul}. It is worth to note that this is not equivalent to operate several independent reservoirs in parallel. Indeed, since the carrier population interacts with all the optical modes, they become coupled (second term in Eq.(\ref{eq:1})) \cite{harkhoe2019delay}. Moreover, additional virtual node interaction would arise from Cross Photon Absorption.\\
\noindent
Another aspect of concern is the memory capacity. Since our device lacks of any delay loop, the memory is provided by carrier relaxation, and is of short term. As shown in Section \ref{sec_principle_of_operation}, the memory is of the order of $\sim3\,\tau_{\textup{fc}}$, which only extends over the first neighboring virtual nodes. Moreover, the terms in Eq.(\ref{eq:1}) indicate that the memory on the past inputs of the pump ($\propto u_p^2$) and of the virtual nodes ($\propto u_{pr}u_p$) is nonlinear. There is a clear unbalance between memory and nonlinearity, the second being preponderant, which is known to be sub-optimal for a certain class of tasks \cite{inubushi2017reservoir}. However, we believe that this is not a severe limitation. It has been already shown and experimentally validated that by providing past inputs to the input layer of a memory-less reservoir, the latter performs as well as the ones with delayed feedback loops. \cite{Memory_Time,Memory_OSA}. 

\section{Conclusions}
\label{sec_conclusion}
In this work, we have experimentally shown a time multiplexed architecture which uses a single integrated MR as a substrate for RC. We encode the input layer in the temporal evolution of the intensity of a pump laser. When this is resonantly coupled to the MR, the information is transferred to a secondary probe laser, which is tuned on an adjacent resonance order. The transfer is enabled by TPA, which generates a free carrier dynamics that is imprinted on the probe laser by FCD. We train the reservoir to perform some proof of concept tasks by using regularized linear regression. We experimentally demonstrate that the system can handle problems with binary and analog inputs. We have shown that the MR solves the 1-bit delayed XOR task with a minimum BER of $1.4\times10^{-3}$ for bitrates up to $30\,\textrm{MHz}$. This result is obtained by only exploiting the nonlinear free carrier dynamics. As a second step, we have shown that the system can solve the analog task of the classification of the Iris dataset. \\
\noindent We demonstrated a maximum classification rate of $(99.3\pm0.2)\%$ at a rate of $0.38\,\textrm{MHz}$, using an average power of $7\,\textrm{dBm}$ at the input of the MR. In this regime, the thermal and the free carrier dynamics are strongly coupled. This feature enhances the separability properties of the reservoir, by adding more variability to the virtual node distribution. The optimal performance is found across the edge between a stable and a self-pulsing regime, which agrees with the predictions of numerical simulations.\\
\noindent
To the best of the author's knowledge, this is the first experimental demonstration of RC using a silicon MR. We envisaged several future developments and engineering solutions, all leveraging on silicon photonics. This will open the doors to the  realization of a wide range of hybrid spatio-temporal reservoirs offering increased computational complexity. 
\section*{Appendix A: Derivation of the reservoir dynamical equation}
Here we derive Eq.(1) of the main text, which describes the time evolution
of the probe at the output of the Drop port of the resonator. The starting
point is the equation governing the free carrier dynamics, which is
given by \cite{johnson2006self}:
\begin{equation}
\frac{d\Delta N}{dt}=-\frac{\Delta N}{\tau_{\textup{fc}}}+g_{\textup{tpa}}U^{2},\label{eq:A1}
\end{equation}
where $g_{\textup{tpa}}$ is the free carrier generation rate per unit
energy induced by TPA and $U$ is the internal energy of the resonator. Equation (\ref{eq:A1})
has the formal solution:
\begin{equation}
\Delta N(t)=g_{\textup{tpa}}\int_{-\infty}^{t}e^{-\frac{t-\xi}{\tau_{\textup{fc}}}}U^{2}(\xi)d\xi.\label{eq:A2}
\end{equation}
The total energy is given by $U=U_{p}+U_{pr}$, i.e., the sum of the
energies of the pump ($U_p$) and probe ($U_{pr}$) lasers. Since the probe is weaker than the pump, we let $U\sim U_{p}$. From temporal coupled mode theory, the expression
for $U_{p}$ is given by $U_{p}=f_{p}P_{p}$, where
$f_{p}$ is defined as \cite{biasi2018hermitian}:
\begin{equation}
f_{p}=\frac{\gamma_{e}}{\left(\omega_{p,0}(1+\delta\omega)-\omega_{p}\right)^{2}+\gamma_{\textup{tot}}^{2}},\label{eq:A3}
\end{equation}
in which $\gamma_{\textup{tot}}$ is the total loss rate of photons
from the cavity, $\gamma_{e}$ the extrinsic loss in the bus waveguide,
and $\delta\omega$ the normalized frequency shift imparted by
thermal and free carrier dispersion. The latter is given by $\delta\omega=-\frac{\Gamma}{n}\left(\sigma_{\textup{FCD}}\Delta N+\sigma_{\textup{TOE}}\Delta T\right)$,
where $\Gamma$ is the modal confinement factor, $n$ the refractive
index of the waveguide core material, $\sigma_{\textup{FCD}}$ the FCD coefficient,
$\sigma_{\textup{TOE}}$ the thermo optic coefficient and $\Delta T$
the differential temperature of the MR with respect to the cold cavity condition. An similar expression holds for $f_{pr}$,
which defines $U_{pr}=f_{pr}P_{pr}$. Note that Eq.(\ref{eq:A3}) assumes that the MR internal energy adiabatically follows the input power variations $P_p$. In our case, this holds since $P_p$ varies on the timescale of the free carrier lifetime \mbox{$\tau_{\textup{fc}}=45\,\textrm{ns}$}, while the photon lifetime is $\frac{2}{\gamma_{\textup{tot}}} \sim  10\,\textrm{ps}$. Moreover, in our experiment the pump power varies much faster than the thermal decay constant of the MR \cite{borghi2020OntheModeling}, so that after an initial transient, the temperature reaches an equilibrium value $\overline{\Delta T}$. As discussed in Section 3.C, for our MR this approximation holds up to $P_{p}=6\,\textrm{dBm}$, after which thermal effects drive the cavity in an unstable regime and have to be accounted for in the dynamics. We now expand  $f_p^2$ to the first order in $\Delta N$ around a reference value $\overline{\Delta N}$  as $f_{p}^{2}\sim \overline{f_{p}^{2}}+\frac{df_{p}^{2}}{d\Delta N}\left(\Delta N-\overline{\Delta N}\right),$
where:
\begin{equation}
\frac{df_{p}^{2}}{d\Delta N}=\frac{4\frac{\Gamma\sigma_{\textup{FCD}}}{n}\omega_{p,0}\gamma_{e}^{2}\left[\left(1+\delta\omega\right)\omega_{p,0}-\omega_{p}\right]^{2}}{\left[\left(\left(1+\delta\omega\right)\omega_{p,0}-\omega_{p}\right)^{2}+\gamma_{\textup{tot}}^{2}\right]^{3}}.\label{eq:A4}
\end{equation}
In deriving Eq.(\ref{eq:A4}), we neglected the dependence of $\gamma_{\textup{tot}}$
on both $\Delta N$ and $U$, which is due respectively to FCA and TPA, since their contribution is small compared to FCD. If the pump power variations around the
reference value $\overline{P}_{p,0}$, giving the equilibrium
condition $\Delta N=\overline{\Delta N}$, are sufficiently small
to justify the first order expansion of $f_p^2$, we can insert it into Eq.(\ref{eq:A2}) to give:
\begin{equation}
\Delta N(t)=g_{\textup{tpa}}\int_{-\infty}^{t}e^{\frac{t-\xi}{\tau_{\textup{fc}}}}P_{p}^{2}(\xi)\left(\overline{f_{p}^{2}}+\frac{df_{p}^{2}}{d\Delta N}\left(\Delta N(\xi)-\overline{\Delta N}\right)\right)d\xi.\label{eq:A5}
\end{equation}
Since the probe and the pump lasers have similar detunings and linewidth $\gamma_{\textup{tot}}^{-1}$ with the corresponding MR resonances, we can also expand $f_{pr}$ to the first order in $\Delta N$.
This allows to relate $\Delta N$ to the probe energy $U_{pr}$
as: 
\begin{equation}
\Delta N-\overline{\Delta N}=\left(P_{pr}\frac{df_{pr}}{d\Delta N}\right)^{-1}\left(U_{pr}-\overline{U_{pr}}\right).\label{eq:A6}
\end{equation}
By substituting Eq.(\ref{eq:A6}) into Eq.(\ref{eq:A5}) and rearranging
terms, we obtain Eq.(1) of the main text, with the definitions:
\begin{equation}
c_{0}=\gamma_e \left (\overline{U_{pr}}-\frac{df_{pr}}{d\Delta N}\overline{\Delta N}P_{pr} \right ),\label{eq:A7}
\end{equation}
\begin{equation}
c_{1}=\frac{df_{pr}}{d\Delta N}g_{\textup{tpa}}\gamma_e\overline{f_{p}^{2}} P_{pr}\left(1-\frac{1}{\overline{f_{p}^{2}}}\left(\frac{df_{p}^{2}}{d\Delta N}\right)\left(\frac{df_{pr}}{d\Delta N}\right)^{-1}\frac{\overline{U_{pr}}}{P_{pr}}\right),\label{eq:A8}
\end{equation}
\begin{equation}
c_{2}=g_{\textup{tpa}}\gamma_e\frac{df_{p}^{2}}{d\Delta N}.\label{eq:A9}
\end{equation}
By taking the values of the varius coefficients from \cite{borghi2020OntheModeling}, and by using $P_{pr} = 0.5\,\textrm{mW}$, one finds $c_1\sim 2430\,\textrm{fJ}^{-1}$ and $c_2\overline{U_{pr}}\sim-c_1$. 
\section*{Appendix B. Relative magnitude between Kerr and Free Carrier effects}
\label{Appendix_B}
Using the same notation of Appendix A, the normalized resonance shift
$\delta\omega$ imparted by Kerr and FCD are given respectively by
$\delta\omega_{\textup{FCD}}=-\frac{\Gamma}{n}\sigma_{\textup{FCD}}\Delta N$
and $\delta\omega_{\textup{Kerr}}=-\frac{cn_{2}U_{p}}{n^{2}V}$ \cite{borghi2016linear},
where $n_{2}$ is the nonlinear refractive index of silicon, $c$
the speed of light and $V$ the volume of the MR. We can calculate
$\Delta N$ as the steady state solution of Eq.(\ref{eq:A1}) as $\Delta N=\tau_{\textup{fc}}g_{\textup{tpa}}U_{p}^{2}$, that
combined to Eq.(\ref{eq:A3}) yields the following expression for
$\delta\omega_{\textup{fc}}$:
\begin{equation}
\delta\omega_{\textup{fc}}=-\frac{\sigma_{\textup{FCD}}\tau_{\textup{fc}}g_{\textup{tpa}}q^{2}(\omega_{p},\omega_{p,0})Q^{2}P_{p}^{2}}{16\omega_{p}^{2}n},\label{eq:A10}
\end{equation}
where $q(\omega_{p},\omega_{p,0})$ is defined as:
\begin{equation}
q(\omega_{p},\omega_{p,0})=\left (\left(\frac{\omega_{p,0}}{\gamma_{\textup{tot}}}\left(1+\delta\omega-\frac{\omega_{p}}{\omega_{p,0}}\right)\right)^{2}+1\right )^{-1}. \label{eq:12}    
\end{equation}
For simplicity, we assumed $\gamma_{\textup{tot}}\sim2\gamma_{e}$,
which in our case is justified by the fact that $Q_{i}=\frac{\omega}{\gamma_{i}}\gg Q_{e}$ (see Section 3 of the main text). Similarly, the expression for $\delta\omega_{\textup{Kerr}}$ is
given by:
\begin{equation}
\delta\omega_{\textup{Kerr}}=-\frac{cn_{2}q(\omega_{p},\omega_{p,0})QP_{p}}{4\omega_{p}n^{2}V}.\label{eq:A11}
\end{equation}
We can now calculate the ratio between the FCD and the Kerr frequency shift as:
\begin{equation}
\frac{\delta\omega_{\textup{fc}}}{\delta\omega_{\textup{Kerr}}}=\frac{\sigma_{\textup{FCD}}\tau_{\textup{fc}}g_{\textup{tpa}}q(\omega_{p},\omega_{p,0})nVQP_{p}}{4\omega_{p}cn_{2}}.\label{eq:A12}
\end{equation}
By inserting the values of the parameters in \cite{borghi2020OntheModeling}, and using $\sigma_{\textup{FCD}}=-4\times10^{-27}\textrm{m}^{3}$
\cite{lin2007nonlinear}, $P_{p}=5\,\textrm{mW}$ and $q=0.95$ (which corresponds to the experimental
pump detuning of $-7.5\,\textrm{GHz})$ we obtain $\frac{\delta\omega_{\textup{fc}}}{\delta\omega_{\textup{Kerr}}}\sim200$.
We expect the actual ratio to be slightly lower due to the
degradation of the quality factor caused by TPA and FCA, which however does
not alter the claim that FCD imparts a resonance shift which is two
orders of magnitude higher than the one of the Kerr.
\section*{Acknowledgements}
The authors acknowledge Dr. Mattia Mancinelli and Mr. Davide Bazzanella for the technical support in the experiment and the fruitful discussions.
\section*{Funding}
This project has received funding from the European Research Council (ERC) under the
European Union’s Horizon 2020 research and innovation programme (grant agreement No
788793, BACKUP), and from the MIUR under the project PRIN PELM (20177 PSCKT).

%


\end{document}